\documentclass[prd,aps,floats,twocolumn,showpacs]{revtex4}
\usepackage[dvips]{epsfig}
\begin{document}
\voffset 15mm
\title{Dynamics of Phase Transitions: The 3D 3-state Potts Model}

\author{ Bernd A. Berg$^{\rm \,a,b}$, Hildegard 
Meyer-Ortmanns$^{\rm \,d}$ and Alexander Velytsky$^{\rm \,a,b}$}

\affiliation{ (E-mails: berg@hep.fsu.edu, 
h.ortmanns@iu-bremen.de, velytski@csit.fsu.edu)\\ 
$^{\rm \,a)}$ Department of Physics, Florida State University,
  Tallahassee, FL 32306\\
$^{\rm \,b)}$ School of Computational Science, 
Florida State University, Tallahassee, FL 32306\\
$^{\rm \,d)}$ School of Engineering and Science, International
University Bremen, P.O. Box 750561, D-28725 Bremen, Germany
 } 

\date{\today }

\begin{abstract}

In studies of the QCD deconfining phase transition or cross-over by 
means of heavy ion experiments, one ought to be concerned about 
non-equilibrium effects due to heating and cooling of the system. In 
this paper we extend our previous study of Glauber dynamics of 2D Potts 
models to the 3D 3-state Potts model, which serves as an effective model 
for some QCD properties. We investigate the linear theory of spinodal 
decomposition in some detail. It describes the early time evolution of 
the 3D model under a quench from the disordered into the ordered phase 
well, but fails in 2D. Further, the quench leads to competing vacuum 
domains, which are difficult to equilibrate, even in the presence of a 
small external magnetic field. From our hysteresis study we find, as 
before, a dynamics dominated by spinodal decomposition. There is
evidence that some effects survive in the case of a cross-over. But 
the infinite volume extrapolation is difficult to control, even with 
lattices as large as $120^3$.


\end{abstract}
\maketitle



\section{Introduction}

In a previous paper~\cite{BHMV}, which we denote by I in the following,
we investigated the dynamics of 2D Potts models in the Glauber 
universality class. Dynamics in this class, model~A in the 
classification of Ref.\cite{ChLu97}, contains local Monte Carlo (MC) 
updating schemes \cite{MC_Glauber} which imitate the thermal fluctuations 
of nature. Here we extend our study to the 3D 3-state Potts model with 
and without an external magnetic field. This model gives an effective 
description of the deconfinement transition \cite{SY82,Gr83,Og84,GoOg85} 
and by adding an external field \cite{BU83}, one can represent the effect 
of quark masses. 

A detailed motivation of our study in the context of QCD and a 
compilation of preliminaries such as Potts models, spinodal 
decomposition, Fortuin-Kasteleyn (FK) clusters and structure functions,
are given in the first two sections of I and are not repeated here.  
Hysteresis calculations for 2D Potts models are presented in I for a 
number of observables including the (internal) energy, properties of FK 
clusters and structure function. The main result is that the dynamics 
leads for most parameter values to a spinodal decomposition, which 
dominates the statistical properties of the configurations, 
independently of whether the equilibrium transition is second 
or first order. Hysteresis cycles and time evolution after a 
cooling \cite{cooling} quench are presented in I and we find also 
initial evidence (compare Fig.~18 of I) that the scenario of dynamical 
spinodal decomposition may survive when the equilibrium transition is 
rendered into a cross-over. The latter point is of importance for QCD,
because lattice simulations and effective models suggest that for 
physical quark masses $m_u$, $m_d$ of the order of 10 MeV and $m_s$ 
of the order of 150 MeV  neither a chiral nor a deconfining transition
occurs in the sense that there are thermodynamical singularities
\cite{crossover}.

In the linear approximation of spinodal decomposition Cahn-Hilliard
theory \cite{CaHi58,Ca68} for model~B predicts an exponential growth 
of the low momentum structure function in the initial part of the time 
evolution.
Whether such an exponential growth was found or not was used by
Miller and Ogilvie \cite{MiOg02} to determine a critical $k_c$ 
between low and high momentum modes, which they related to the 
Debye screening mass. In this paper we present the linear 
approximation, derive its diffusive differential equation,
and confront the predicted exponential growth with MC data.

At the present level our results are mainly qualitative. Within a model
of dynamics one expects universal properties, because the equations 
of motion differ only in their constants, but not in their general 
structure. The proper dynamics for Minkowskian relativistic dynamics
is expected to be hyperbolic. Altogether little is known about 
the kind of effects one may expect to be produced by a dynamics. 
Therefore, we would like to advocate studies of different models of 
dynamics to get a handle on their similarities and differences. We 
begin with the diffusive dynamics of model~A by the pragmatic reason
that it is relatively easy to follow the time evolution of corresponding
MC algorithms. The study of hyperbolic dynamics within full QCD or just
pure non-Abelian gauge theory is presently out of reach. However the
Polyakov loop model of Pisarski and Dumitru \cite{DuPi01} allows for
simulations in the Minkowskian \cite{Sc01} and it may be interesting 
to confront an investigation of its dynamics with our present results.

In the next section we fill a gap of the literature by presenting the
linear theory of spinodal decomposition for model~A. Our numerical
results are compiled in section~\ref{sec_nr}, with quench and 
hysteresis investigations in subsections. Summary and conclusions 
are given in the final section~\ref{sec_conclude}.
\vfil

\section{Spinodal Decomposition in the Linear Approximation 
\label{sec_spinodal}}

Metastable states are non-equilibrium states with lifetimes comparable 
to the time of observation, while unstable states are non-equilibrium 
evolving states with no observable lifetime \cite{GuDr85,La92}. When 
dealing with metastable states we consider systems with a first order 
phase transition, while unstable states can be observed for systems 
with first and second order phase transitions.

Metastable and unstable states are distinguished by different types 
of instability characterizing the early stages of a phase separation.
Metastable states require finite amplitude, localized fluctuations. 
The system evolves by growing the nuclei or droplets of the new phase 
and the theory of nucleation explains (e.g. Becker-D\"oring theory
\cite{GuDr85,La92}) the rate of formation of such nucleating droplets. 
In contrast, infinitesimal amplitude, non-localized fluctuations are 
sufficient for the initial decay of unstable states, which is called 
spinodal decomposition.

In mean field theory there is a sharp distinction between metastable
and unstable states. They are separated by the spinodal curve, which is 
the locus of points inside the coexistence curve for which an appropriate
susceptibility (for instance the concentration susceptibility 
$\partial c/\partial \mu$ for binary fluids) diverges. There is 
no sharp distinction in real systems.

There is a difference in the dynamics for models with conserved 
(model~B in the classification of \cite{ChLu97}) and non-conserved 
(model~A) order parameters. Binary alloy or binary fluid models are 
typical examples of models with conserved order parameters (the 
concentration $c$). Originally \cite{CaHi58,Ca68} the linear 
approximation was developed for model~B. Following \cite{Vel} we 
consider here a linear theory of spinodal decomposition for model~A. 
Related results can also be found in \cite{BrRi02}. The general picture 
is the same for model~A and model~B dynamics. The difference is in the
quantitative details.

\subsection{Coarse-grained free energy}

We start by considering the coarse-grained free energy \cite{CaHi58}. 
It is a functional of the order parameter and its minima correspond to 
the equilibrium states of the model. To use the coarse-grained energy 
for a study of non-equilibrium phenomena, one assumes that the model is 
not too far from equilibrium. The requirement is fulfilled for systems 
with slow dynamics on the scale of local equilibration times.

One introduces a spatial mesoscopic length scale $l$, separating 
the fast dynamics local microscopic regions from the slow dynamics 
macroscopic regions. The system is divided into regions ({\em elementary 
physical cells}\index{elementary physical cell}) of size $l$, which are 
centered around vector coordinates $\vec{r}$ defining their positions. 
The length scale $l$ has to be significantly larger than the 
lattice spacing size, $l>>a$, to allow for a statistical treatment of 
the elementary physical cells. The other requirement is that $l\ll \xi$, 
where $\xi$ is a typical correlation length of the system. On the scale 
$l$ one treats the system as being in the equilibrium. The order 
parameter $m(\vec{r},t)$ (e.g. the magnetization) has to vary smoothly 
on the length scales larger than $l$, to satisfy the requirement that 
the system on these length scales is not far from equilibrium. 

Since elementary physical cells are equilibrated, their equilibrium 
thermodynamical functions are defined. We write down the coarse-grained
partition function for the system of cells
\begin{equation}
Z_l(\{m\})=\sum^\prime_{\{S\}}\exp(-E\{S\}/k_BT),
\label{coarse_gr_z}
\end{equation}
where the sum is over spin configurations $\{S\}$ and the prime indicates the
constraint
\begin{equation}
\frac1{l^d}\sum_{\{S\}\,\in\,\mbox{cell }\vec{r_i}} S_i=m(\vec{r_i}).
\end{equation}
By summing in Eq.~(\ref{coarse_gr_z}) over all configurations $\{m\}$ of 
magnetization, one gets the normal partition function.

The coarse-grained free energy is
\begin{equation}
F_l(\{m\})=-k_BT\ln Z_l(\{m\})=
\int d^dr f(m(\vec{r})),
\end{equation}
where we introduced the non-uniform free energy density $f(m(\vec{r}))$ and
treat the magnetization as a continuous function of $\vec{r}$.
We assume that $f$ is a function of the local magnetization and
its derivatives and expand it about $f_0$, the free energy per spin
of a uniform phase. For a cubic lattice $f(m,\nabla m,\dots)=f_0(m)+k_1
\nabla^2m+k_2(\nabla m)^2+\dots\,$ and
\begin{equation}
F_l(\{m\})\simeq\int d^dr[f_0(m)+k_1\nabla^2m+k_2(\nabla m)^2].
\end{equation}
Applying Green's first identity \cite{Jackson2} and imposing that 
$\nabla m$ is zero at the boundary, we obtain
\begin{eqnarray}
\int d^dr\, k_1 \nabla^2 m &=& -\int d^dr\, \nabla k_1 \nabla m\\
&=& -\int d^dr\, \left( \frac{d k_1}{d m} \right)\,
                 \left( \nabla m\right)^2\ . \nonumber
\end{eqnarray}
The coarse-grained free energy can be written as
\begin{eqnarray} \label{gl-free}
F_l(\{m\})=\int d^dr\left[\frac12K(\nabla m)^2+f_0(m)\right],\\
\label{K} {\rm where}~~~~{1\over 2}\,K\ =\ k_2 - {d k_1\over d m}\,. 
\end{eqnarray}
It has the familiar Ginzburg-Landau form. 
Note that $f_0$ depends on the choice of the spatial length scale $l$.

\subsection{The linear approximation}

The phenomenological model we use is based on the assumption that 
the rate of displacement of the order parameter is linearly proportional 
to the local thermodynamic force $\delta F/\delta m(\vec{r})$
\begin{equation}
\frac{\partial m}{\partial t}=-\Gamma\frac{\delta F}{\delta m(\vec{r})}.
\end{equation}
The coefficient of the proportionality, $\Gamma$, is the response 
coefficient and defines the relaxation time scale of the system. This 
equation describes a purely dissipative system and not one with an 
internal Hamiltonian dynamics. 

Let us determine the variation of the coarse-grained free energy 
(\ref{gl-free}) 
\begin{equation}
\delta F_l(\{m\})=\int d\vec{r}\left[\frac12K
\frac{d (\nabla m)^2}{d (\nabla m)}\delta\nabla m+\frac{d f_0(m)}{dm}
\delta m\right]\,.
\end{equation}
Using again Green's first identity and dropping the surface integral, 
we write the variation as
\begin{equation}
\delta F_l(\{m\}) = \int d\vec{r}\, 
\left[-K\nabla^2 m + \frac{d f_0(m)}{dm}\right] \delta m\,.
\end{equation}
The functional derivative is 
\begin{equation}
\frac{\delta F}{\delta m(\vec{r})}=-K\nabla^2 m+\frac{d f_0(m)}{dm}
\end{equation}
and the equation of motion becomes
\begin{equation}
  \frac{\partial m}{\partial t}=
  -\Gamma\left[-K\nabla^2m+\frac{df_0}{dm}\right]\,.
\end{equation}
The Langevin approach \cite{La92} gives the same equation, but with 
a noise term added. We are interested in the fluctuations of the 
magnetization about some average value $m_0$
\begin{equation}
  m(\vec{r},t)\,=\,m_0+u(\vec{r},t)\,.
\end{equation}
To solve the differential equation we write it for small fluctuations 
\begin{equation} \label{deqn}
  \frac{\partial u(\vec{r},t)}{\partial t} \simeq 
\end{equation}
$$ -\Gamma\left[-K\nabla^2u(\vec{r},t)+\frac{df_0(m_0)}{dm_0}+
    u(\vec{r},t)\,\frac{d^2f_0(m_0)}{dm_0^2}\right] $$
and represent $u(\vec{r},t)$ as a Fourier series
\begin{equation}
  u(\vec{r},t)=
  \frac1V\sum_{\vec{k}}\hat{u}(\vec{k},t)\exp(i\vec{k}\cdot\vec{r})\ .
\end{equation}
Using the Fourier series Eq.~(\ref{deqn}) becomes
\begin{eqnarray} \nonumber
  \frac1V\sum_{\vec{k}} \left[
  \frac{\partial\hat{u}(\vec{k},t)}{\partial t} - \omega(\vec{k})\,
  \hat{u}(\vec{k},t) \right] \exp(i\vec{k}\cdot\vec{r}) \\
   =\, -\Gamma\frac{df_0(m_0)}{dm_0}\qquad\qquad\qquad \\ 
\label{omega_k} {\rm where}~~~~\omega(\vec{k})\,
  =\, -\Gamma\left(Kk^2+\frac{d^2f_0(m_0)}{dm_0^2}\right)\ .
\end{eqnarray}
Multiplying both sides by $\sum_{\vec{r}}\exp
(-i\vec{k}^\prime\cdot \vec{r})$
and using the Kronecker function relation 
$$V\delta(\vec{k}-\vec{k}^\prime) = 
  \sum_{\vec{r}}\exp\left(i(\vec{k}-\vec{k}^\prime)\cdot\vec{r}\right)$$ 
we get
\begin{eqnarray} \label{u_eq_mot}
  \frac{\partial \hat{u}(\vec{k}^\prime,t)}{\partial t} &-& 
  \omega(\vec{k}^\prime)\,\hat{u}(\vec{k}^\prime,t)\, =\, g(k^\prime)\\
  {\rm where}~~~~ g(k) &=& - \Gamma\frac{df_0(m_0)}{dm_0}
  \sum_{\vec{r}}\exp(i\vec{k}\cdot \vec{r})\ .
\end{eqnarray}
The general solution is 
\begin{equation}
  \hat{u}(\vec{k}^\prime,t)=C\exp\left(\omega(\vec{k}^\prime)\,t\right)
  -\frac{g(\vec{k}^\prime)}{\omega(\vec{k}^\prime)}\,.
\end{equation}
Except for the linear term this equation is the same as the one produced
for model~B by Cahn-Hilliard theory \cite{CaHi58} and a similar analysis 
is appropriate. If 
$$ \frac{d^2f_0(m_0)}{dm_0^2} > 0 $$ 
holds, Eq.~(\ref{omega_k}) implies that the amplitude of any fluctuation 
approaches a constant exponentially fast with time. But if the second 
derivative is negative, then one sees an exponential growth of the 
fluctuations for momentum modes smaller than the critical value 
$$ k < k_c = 
   \left[-\frac1K \frac{d^2f_0(m_0)}{dm_0^2}\right]^{1/2}\,. $$ 

\subsection{Equation of motion for the structure factor\label{str_fact1}}

The structure factor (or function) can be measured in condensed matter 
experiments: A system is exposed to radiation of a wavelength $\lambda$ 
and the scattering intensity at different angles $\theta$ is recorded. 
The structure factor $\hat{S}(\vec{k},t)$ is proportional to the 
scattering intensity at the angle $\theta$, where $k = (4\pi/\lambda)
\sin(\frac12\theta)$. 
In the gauge theory studies one may expect that the structure factor 
reflects the production of corresponding momentum gluons.
The structure function can be written as 
\begin{eqnarray} \nonumber
  \hat{S}(\vec{k},t) &=& \int \left\langle u(\vec{r},t)\,
  u(\vec{r^\prime},t) \right\rangle \exp\left(i\vec{k}\cdot(\vec{r}
  - \vec{r^\prime})\right)\,d\vec{r}\,d\vec{r^\prime} \\
  &=& \left\langle\hat{u}(\vec{k},t)\,\hat{u}(-\vec{k},t)
      \right\rangle \label{str_f1}
\end{eqnarray}
To derive the equation of motion for the structure factor we take the 
time derivative of $\hat{S}(\vec{k},t)$ and make use of 
Eq.~(\ref{u_eq_mot})
\begin{eqnarray} \nonumber
  \frac{\partial \hat{S}(\vec{k},t)}{\partial t} &=& \left\langle 
  2\,\hat{u}(\vec{k},t)\,\hat{u}^*(\vec{k},t)\,\omega(k)\right\rangle \\
  &+& \left\langle \left(\hat{u}^*(\vec{k},t)+\hat{u}(\vec{k},t)
      \right) g(k) \right\rangle\,. \label{s_eq_mot1}
\end{eqnarray}
The average of fluctuations about the average magnetization has to be 
zero $\left\langle\hat{u}(\vec{k},t)\right\rangle=0$. Thus 
(\ref{s_eq_mot1}) becomes
\begin{equation}
  \frac{\partial \hat{S}(\vec{k},t)}{\partial t}
  = 2\,\omega(\vec{k})\,\hat{S}(\vec{k},t)\,,
\end{equation}
with the solution
\begin{equation} \label{str_fact}
  \hat{S}(\vec{k},t)=
  \hat{S}(\vec{k},t=0)\exp\left(2\omega(\vec{k})t\right)\,.
\end{equation}
We see that it is very similar to the evolution of fluctuations.
Again, if  $d^2f_0(m_0)/dm_0^2<0$ low momentum modes grow exponentially. 
The value of the critical momentum is the same as for the fluctuations. 
The maximum growing mode is the one for which $\omega(\vec{k})$ 
takes on its maximum, $\omega_{\max}=\omega(k_{\max})$. In the linear
theory $k_{\max}=0$ according to Eq.~(\ref{omega_k}). The values of 
$\omega (\vec{k})$ decrease with increasing $k$. Then there exists a 
$k_c$ value, so that higher momentum modes $k>k_c$ are exponentially 
decaying. The results in the linear approximation are the same as for 
models with conserved order parameter, except for the 
values of $\omega(\vec{k})$ and $k_{max}$.

A similar structure factor behavior may be obtained from the somewhat
more accurate Langevin approach if we consider only linear terms. This 
result was first obtained for model~B by H. Cook according to the 
review~\cite{La92}. For model~A a straightforward algebra yields
\begin{equation} \label{langevin_sf}
  \frac{\partial \hat{S}(\vec{k},t)}{\partial t} =
  2\omega(\vec{k})\hat{S}(\vec{k},t)+2k_BT\Gamma+2g(k)\,.
\end{equation}
When the intrinsic dynamics of the system is prevailing over the 
extrinsic thermal fluctuations than the right hand side is dominated 
by the first term and the previous result for the structure function 
(\ref{str_fact}) is recovered.

\subsection{Validity of the linear approximation\label{linear_appr}}

In the following we assume the existence of two phases, each described 
by a single equilibrium value of an order parameter. Potts models belong 
to a slightly more general group, because there are $q=2,\,3,\dots$
equilibrium values in the ordered phase. To trace them in our present 
discussion would lead to a rather cumbersome notation. Now
the uniform free energy $f_0$ at the scale $l$ has two minima which 
correspond to the equilibrium values of the order parameter in the two 
phases: $m_1$ and $m_2$. The obvious criterium for the validity of the 
linear approximation is that the mean square value of fluctuations of 
the magnetization are much smaller than the square of the magnetization 
scale, $\Delta m=m_2-m_1$: $\langle u^2\rangle<<(\Delta m)^2$. 

Note that the fluctuations $u$ depend on the introduced coarse-grained 
scale. The fluctuations are averaged over the cells of volume 
$l^d\sim k^{-d}_{c}$
\begin{equation}
  \left\langle u^2(\vec{r},t)\right\rangle = S(0,t) =
  \int d\vec{k}\,\hat{S}(\vec{k},t) \sim k^d_{c}\hat{S}(0,t)
\end{equation}
and we can write the condition for the validity of the linear 
approximation as
\begin{equation}
  k^d_{c}\hat{S}(0,t) \ll (\Delta m)^2.
\end{equation}
The Langevin approach allows us to estimate the value of $\hat{S}(0,t)$ 
from the requirement that the first term of the right-hand side of 
(\ref{langevin_sf}) is the largest (see \cite{La92} for similar
consideration for model~B) 
and gives
\begin{equation} \label{linear1}
  \frac{k_BTk^{d}_{c}}{\left|d^2f_0(m_0)/dm_0^2\right|(\Delta m)^2}
  =\frac{k_BTk^{d-2}_{c}}{K(\Delta m)^2} \ll 1\,.
\end{equation}
If we assume that the system is near the critical point $T_c$, this
equation is {\em not} fulfilled. Close to $T_c$ the only relevant length 
scale is the correlation length $\xi\sim k^{-1}_{c}$. The analysis of 
planar interfaces \cite{La92} shows that $K(\Delta m)^2/\xi$ is 
essentially a surface tension $\sigma$ in a scaling sense. So we can 
rewrite (\ref{linear1}) as
\begin{equation}
  \frac{k_BT_c}{\sigma\xi^{d-1}} \ll 1\,.
\end{equation}
But critical point scaling arguments show that the ratio
$(k_BT_c)/(\sigma\xi^{d-1})$ is a universal constant of order unity 
for $d<4$. This means that for near-critical quenches the linear 
approximation is not valid and non-linear behavior is relevant.

For quenches far from the transition point $T\ll T_c$ the mean-field 
approach becomes accurate. It allows to evaluate the condition 
(\ref{linear1}) of validity of the linear approximation. Using 
mean-field estimates one finds inequality (\ref{linear1}) to hold 
for both model~A and B \cite{La92a}:
\begin{equation} \label{linear2}
  \left(\frac{a}{\xi_0}\right)^d\left(\frac{T}{T_c}\right)^2
  \left(1-\frac{T}{T_c}\right)^{d/2-2} \ll 1\,,
\end{equation}
where $a$ is the lattice spacing and $\xi_0$ is the range of the 
interaction. For Potts models $a/\xi_0=1$. Eq.~(\ref{linear2}) is 
satisfied for $d<4$ and $T\ll T_c$. Note the dependence on the number 
of dimensions in both limits considered. 

\section{Numerical Results\label{sec_nr}}

Our Boltzmann weights are $\exp (-\beta E)$ with the energy function
\begin{equation} \label{energy}
  E = - 2 \sum_{\langle \vec{r},\vec{r}'\rangle} 
  \delta_{\sigma(\vec{r},t),\,\sigma(\vec{r}',t)} -
  2h/\beta \sum_{\vec{r}} \delta_{\sigma(\vec{r},t),\,\sigma_0}\,.
\end{equation}
The first sum runs over all nearest neighbor sites $\vec{r}$ and 
$\vec{r}{\,'}$, and $\sigma$ takes the values $1,\dots\,,q$. In this 
paper we rely on symmetric lattices of $N=L^d$ spins in  
$d=2$ and $d=3$ dimensions.

Of the 3D zero magnetic field Potts models the $q=2$ Ising model 
exhibits a second order phase transition, and the transitions are 
first order for $q\ge3$. The model of our interest is the 3D 3-state
model, which in zero magnetic field has a weak first-order phase 
transition at $\beta_{c} = 0.275283\,(6)$. This transition persists 
for small values of the magnetic field $h$ and has a second order 
end-point at
\begin{equation}
  \beta_{c}=0.27469\,(1)\quad {\rm and} \quad h_{c}=0.000388\,(5)
\label{3dpotts_cp}
\end{equation}
as determined in Ref.\cite{KaSt00} (there is a factor of two
difference between our notation here and this reference).

The structure factor is (see I)
\begin{equation} \label{sfk}
S(\vec{k},t) = \frac1{N_s^2}\sum_{q_0=0}^{q-1}
\left\langle\left|\sum_{\vec{r}}\delta_{\sigma(\vec{r},t),q_0}
\exp[i\vec{k}\vec{r}]\right|^2\right\rangle\ .
\end{equation}
Spinodal decomposition is characterized by an explosive growth in 
the low momentum modes, while the high momentum modes relax to their 
equilibrium values.

During our simulations the structure functions are averaged over 
rotationally equivalent momenta and the notation $S_{k_i}$ is used
to label structure functions of momentum $|\vec{k}|=k_i$ where
\begin{equation} \label{momenta}
\vec{k} = {2\pi\over L}\,\vec{n}\ .
\end{equation}
In 2D we recorded the structure function for the modes:
$n_1$: $(1,0)$ and $(0,1)$, 
$n_2$: $(1,1)$, 
$n_3$: $(2,0)$ and $(0,2)$, 
$n_4$: $(2,1)$ and $(1,2)$, 
$n_5$: $(2,2)$. 

In 3D we recorded the modes (including the permutations)
$n_1$: $(1,0,0)$, 
$n_2$: $(1,1,0)$, 
$n_3$: $(1,1,1)$, 
$n_4$: $(2,0,0)$,
$n_5$: $(2,1,0)$,
$n_6$: $(2,1,1)$,
$n_7$: $(2,2,0)$,
$n_8$: $(2,2,1)$ and $(3,0,0)$,
$n_9$: $(3,1,0)$, 
$n_{10}$: $(3,1,1)$,
$n_{11}$: $(2,2,2)$,
$n_{12}$: $(3,2,0)$, 
$n_{13}$: $(3,2,1)$,
$n_{14}$: $(3,2,2)$,
$n_{15}$: $(3,3,0)$,
$n_{16}$: $(3,3,1)$,
$n_{17}$: $(3,3,2)$,
$n_{18}$: $(3,3,3)$.
Note the accidental degeneracy in length for $n_8$.

As in I we measured various properties of FK and geometrical clusters. 
See the definitions given in~I.

For most of our simulations we use the heat bath algorithm with systematic
updating, because we found previously that systematic updating is faster
than random updating. However, for the linear theory of spinodal 
decomposition, one may be interested in the very early time development. 
Then it is advantageous to be able to take data within one sweep and we 
use random updating for a few cases.

\subsection{Quench\label{sec_quench}}

We study the time evolution after a quench from an initial to a final 
temperature 
\begin{equation} \label{T_quench}
  T_{\rm initial}\ \to T_{\rm final} 
\end{equation}
and are interested in the limit of large lattices, $L\to\infty$. To
get a statistically meaningful sample, we performed between 
$20\times 20$ and $32\times 10$ repetitions of each quench. Here the
first number counts the PCs we used in parallel and the second number 
gives the repetitions performed on each PC.

The large number of repetitions required and the various adjustable 
parameters ($T_{\rm initial}$, $T_{\rm final}$ and $L$) make a thorough 
investigation laborious and quite computer time consuming. Therefore, 
we were not able to arrive at conclusion for all questions, which deemed 
interesting to us.

\subsubsection{Structure Functions\label{Qsec_sf}}

Miller and Ogilvie~\cite{MiOg02} investigated the dynamics of SU(2) 
and SU(3) gauge theories after quenching from a low to a high physical 
temperature (corresponding to the $\beta_{\min}\to\beta_{\max}$
in the spin system). They report a critical value $k_c$, so that 
modes grow (do not grow) exponentially for $k<k_c$ ($k>k_c$). Using
an effective potential approach, they were able to relate $k_c$ to the 
Debye screening mass.

\begin{figure}[ht] \vspace{-2mm} \begin{center}
\epsfig{figure=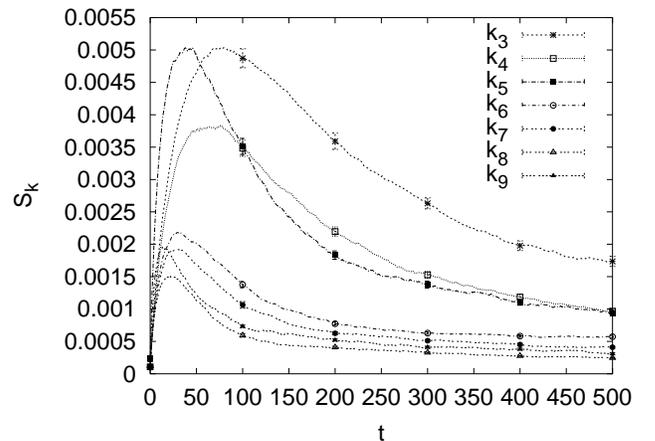,width=\columnwidth} \vspace{-1mm}
\caption{Structure functions for the 2D Ising model quench
$\beta=0.2\to 0.6$ at zero field on an $80\times 80$ lattice.} 
\label{fig_hsfk2} \end{center} \vspace{-3mm} \end{figure}

\begin{figure}[ht] \vspace{-2mm} \begin{center}
\epsfig{figure=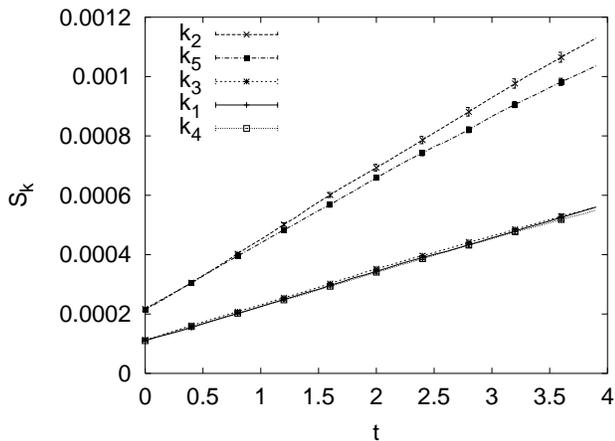,width=\columnwidth} \vspace{-1mm}
\caption{Very early time evolution of the 2D Ising model quench
$\beta=0.2\to 0.6$ at zero field on an $80\times 80$ lattice.
The order in the legend agrees with the order of the curves and 
the lower three curves fall almost on top of one another. } 
\label{fig_hsfk_a} \end{center} \vspace{-3mm} \end{figure}

First we give an example that our 2D Potts model data do not support 
the initial exponential growth predicted by the linear approximation.
Fig.~\ref{fig_hsfk2} depicts the $k\ge k_3$ structure function for 
the 2D Ising model quench $\beta=0.2\to 0.6$ at zero field on an 
$80\times 80$ lattice (the modes $k_1$ and $k_2$ are omitted, because
their peaks are more than four times higher than those of $k_3$ and
$k_4$). As these functions do not show an initial exponential 
increase, random updating was subsequently used to be able to follow
the time development within one sweep. Figure~\ref{fig_hsfk_a} shows
the thus obtained very early time development of the structure functions 
of $S_{k_i}$ for $i=1,\,2,\,3$ and~4. The increase is linear in $t$. The 
curves turn then concave, excluding any exponential growth.

\begin{figure}[ht] \vspace{-2mm} \begin{center}
\epsfig{figure=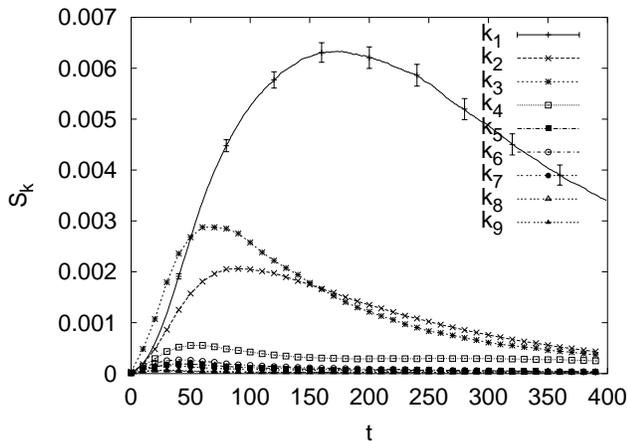,width=\columnwidth} 
\vspace{-1mm}
\caption{Structure functions for the $\beta=0.2\rightarrow0.3$ 
3D 3-state Potts model quench at zero field on a $40^3$ lattice.}
\label{fig_hsfkq03}
\end{center} \vspace{-3mm} \end{figure}

\begin{figure}[ht] \vspace{-2mm} \begin{center}
\epsfig{figure=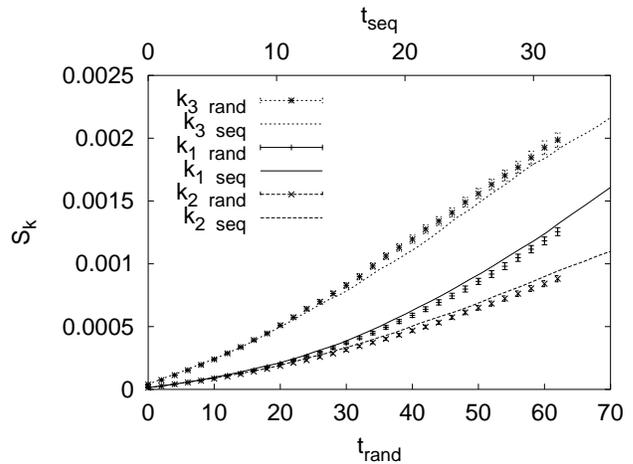,width=\columnwidth}\vspace{-1mm}
\caption{The early time evolution for the quench of
Fig.~\ref{fig_hsfkq03} together with random updating data.
The order in the legend agrees with the order of the curves.
 } \label{fig_hsfkq03_in}
\end{center} \vspace{-3mm} \end{figure}

In Fig.~\ref{fig_hsfkq03} we plot our structure functions for a
$\beta=0.2\to0.3$ quench with sequential updating for the 3D 3-state 
Potts model. Their early time development allows for exponential fits 
of the form~(\ref{str_fact})
\begin{equation} \label{omega_fit}
  C_1(k)+C_2(k)\exp(2\omega(k)t)\,.
\end{equation}
In Fig.~\ref{fig_hsfkq03_in} we show the corresponding random and
sequential updating data (the latter without error bars, not to 
overload the figure). The time between the updating schemes is scaled 
according to $t_{\rm rand} = 1.94\,t_{\rm seq}$.

The fit values for $\omega(k)$ are positive for the low values of $k_i$, 
$i\ge 1$. Notably, the largest value $\omega_{\max}$ is obtained for
$k_3$ in contrast to the monotonous decrease of $\omega(k_i)$ which
the linear theory (\ref{omega_k}) predicts for increasing $k_i$. Already
our 2D results of Fig.~\ref{fig_hsfk_a} disagreed with the ordering of
$\omega(k_i)$
predicted by the linear theory, but there it comes to no surprise, as 
the linear theory fails altogether in 2D. In 3D the $k_3$ mode appears
to be the only major exception, amazingly already observed as such in 
pure gauge theory \cite{MiOg02}, from a picture which is over-all 
consistent with the predicted decrease of $\omega(k_i)$ for increasing 
$k_i$. The fits become quite unstable when $\omega(k)$ approaches zero. 
For even higher $k_i$ values negative $\omega(k)$ values are obtained, 
indicating an initial, exponential approach towards the constant
$C_1(k)$, reaching for larger times a maximum and converting into a
decrease on an always concave curve.

\begin{figure}[ht] \vspace{-2mm} \begin{center}
\epsfig{figure=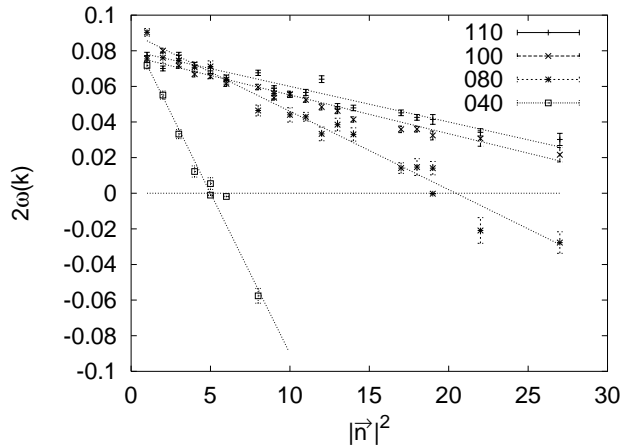,width=\columnwidth} \vspace{-1mm}
\caption{Determination of $k_c$ for the 3D Potts model at
zero field. } \label{fig_omega_3dsq} 
\end{center} \vspace{-3mm} \end{figure}

For sequential updating we have pushed our analysis to $L^3$ lattices 
as large as $L=110$.
The results of all our $\omega (k)$ fits (\ref{omega_fit}) are 
compiled in Fig.~\ref{fig_omega_3dsq} versus $|\vec{n}|^2$ defined
by equation (\ref{momenta}). Approximately, we find straight lines
$\omega(k)=a_0+a_1\,|\vec{n}|^2$ with a negative slope $a_1$ and we 
determine the critical momentum $k_c$ as the value where $\omega(k)$ 
changes its sign. Using (\ref{momenta}) we
find $k_c\approx k_5 = 0.351$ ($L=40$) and $k_c\approx k_{16} =
0.342$ ($L=80$). So the finite size correction from $L=40$ to 
$L=80$ is about 3\%. For $L=100$ and 110 we estimate 
$|\vec{n}_c|^2\ge 31$, outside the range where we took data.
The $k_c$ estimate of \cite{MiOg02} (the Polyakov loops substituting 
for the Potts spins) does not consider fnite size corrections, which 
future studies may reveal.

\subsubsection{FK Clusters\label{Qsec_FK}}

A quench in the temperature changes instantaneously the bond probability 
of FK cluster configurations and observables such as the number of 
clusters, the mean volume and surface, the maximum volume, and maximum 
surface change considerably. Then the evolution proceeds smoothly.

The FK Potts clusters are similar to domains in a ferromagnet. 
For each of the possible magnetizations we define the largest cluster. 
When we quench into the ordered phase one magnetization will eventually 
take over. However, a-priori the system does not know which direction 
this is, because it is prepared in the disordered state where all 
directions are equally probable. Let us address the question how the 
system grows domains. There are two alternatives. (i)~The system grows 
a cluster of a particular magnetization, while clusters of the other 
magnetizations decay right from the beginning. (ii) The system grows 
clusters in each direction of magnetization. At a later stage smaller 
clusters coalesce by statistical fluctuations and one of the directions 
emerges as dominant.

\begin{figure}[ht] \vspace{-2mm} \begin{center}
\epsfig{figure=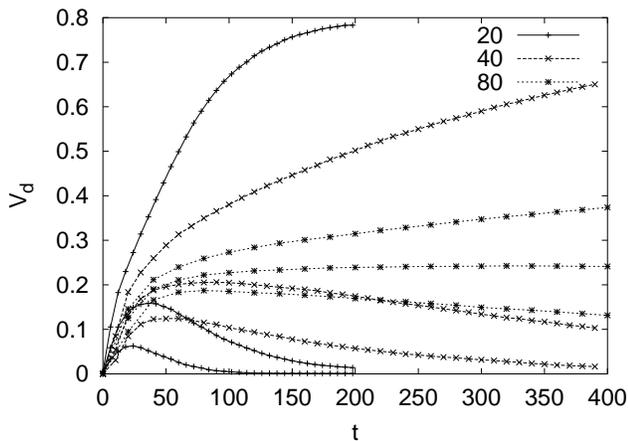,width=\columnwidth} \vspace{-1mm}
\caption{Largest FK clusters of distinct magnetizations for
the 3D 3-state Potts model quench $\beta=0.2\rightarrow0.3$
at zero field on various size lattices.} \label{quench:3qdom}
\end{center} \vspace{-3mm} \end{figure}

In Fig. \ref{quench:3qdom} we plot the evolution of the largest FK 
clusters for the three magnetizations of the 3-state Potts model in 
zero external magnetic field. For different lattice sizes the plot 
shows that the system grows clusters of each magnetization before one 
becomes dominant. The process of competitions between the largest 
clusters of different magnetization takes longer on the larger lattices.

\begin{figure}[ht] \vspace{-2mm} \begin{center}
\epsfig{figure=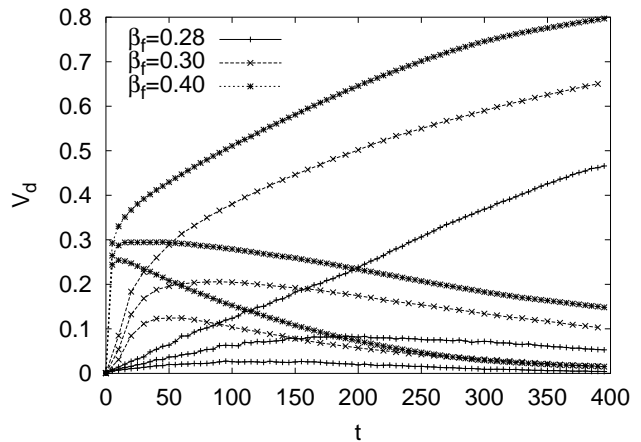,width=\columnwidth} \vspace{-1mm}
\caption{Largest FK clusters of distinct magnetizations for
the 3D 3-state Potts model quenches from $\beta=0.2$ to 
$\beta_f$ at zero external magnetic field on a $40^3$ lattice.} 
\label{quench:3q_diff}
\end{center} \vspace{-3mm} \end{figure}

To study the dependence of the speed of the evolution on the depth of 
the quench, we consider different final temperatures. In 
Fig.~\ref{quench:3q_diff} we compile results for a $40^3$ 
lattice and find that the growth of the largest clusters of all 
the magnetizations is faster when the system is quenched deeper 
into the ordered phase. 

\begin{figure}[ht] \vspace{-2mm} \begin{center}
\epsfig{figure=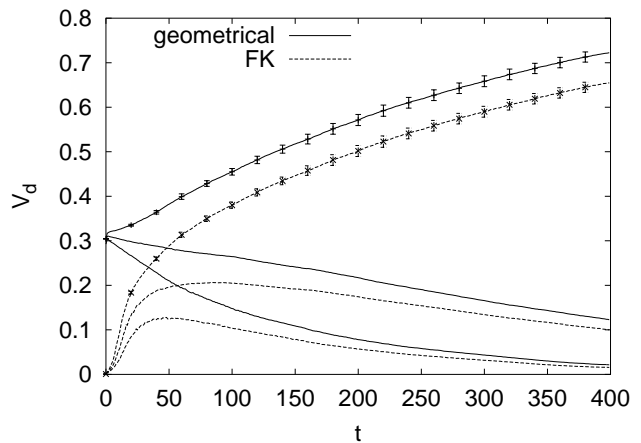,width=\columnwidth} 
\vspace{-1mm}
\caption{Largest geometrical and FK clusters for the 3D 3-state Potts 
model quenched from $\beta=0.2$ to $\beta_f=0.3$ at zero field on a 
$40^3$ lattice.} \label{quench:3q_geom}
\end{center} \vspace{-3mm} \end{figure}

For gauge theories an equivalent percolation model does not exist and 
it is problematic to find an appropriate substitute for the FK 
definition when clusters of Polyakov loops are considered. But it 
is easy to study geometrical clusters. Therefore, we compare for the 
3D 3-state Potts model the evolution of geometrical and FK clusters in 
Fig.~\ref{quench:3q_geom}. The geometrical definition (bonds between
neighboring spins of the same direction are always set) leads to the 
first of our two scenarios. Geometrical clusters do not compete. The 
system starts growing one of the domains, while reducing the largest 
domains of the other magnetizations right after the quench. This 
picture is unfavorable for the use of geometrical clusters of Polyakov 
loops in gauge theories. FK clusters are subsets of geometrical 
clusters. So, the FK clusters of the figure are always smaller than 
the geometrical. Sufficiently far in the ordered phase FK and 
geometrical domains converge in size, while closer to the disordered 
phase the difference is significant, because geometrical clusters lead 
to artificial groupings of spins into clusters. 

\begin{figure}[ht] \vspace{-2mm} \begin{center}
\epsfig{figure=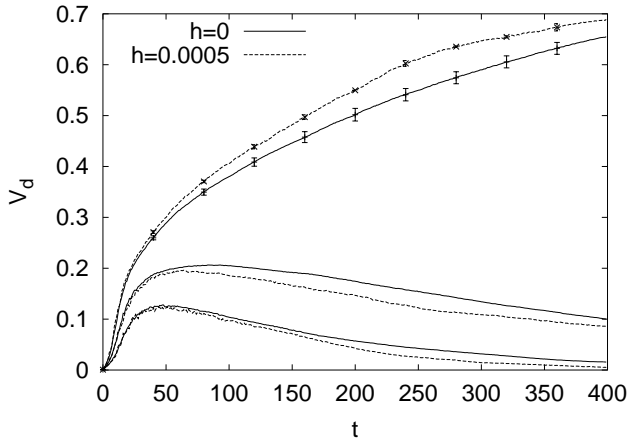,width=\columnwidth} \vspace{-1mm}
\caption{The $3$-D 3-state Potts model quenched from $\beta=0.2$ to 
$\beta_f=0.3$ at zero and $h=0.0005$ external magnetic fields on a
$40^3$ lattice.} \label{quench:3qh}
\end{center} \vspace{-3mm} \end{figure}

Next, we apply the external magnetic field $h=0.0005$, which puts the 
model slightly above the critical endpoint at $h_c=0.000388\,(05)$ in 
the $h-T$ plane, see Eq.~(\ref{3dpotts_cp}), and drive the system 
through the emerging region of a sharp crossover. The external magnetic 
field sets the direction of preferable magnetization. In
Fig.~\ref{quench:3qh} we see that the 
growth proceeds in the same manner as for $h=0$ with FK clusters of all 
magnetizations growing at the early time, only that the preferred 
domain grows a bit faster than before. 
This indicates that this effect may to some extent survive in QCD 
studies of the crossover region and finite size behavior in the
$L\to\infty$ limit ought to be studied, although the physical volume
relevant for RHIC experiments can only be approximated withing lattice
gauge theory.

\subsection{Hysteresis} \label{subsec_cluster}

In our 3D hysteresis investigation we look at the first order phase 
transition and rapid crossover region of the 3-state Potts model. We 
no longer study the limit of very slow dynamics (Eq.~(7) of~I), but use 
volume independent step-sizes
\begin{equation} \label{step_size}
  \triangle\beta^\prime = 
  \frac{2(\beta_{\max}-\beta_{\min})}{n'_{\beta}\, L^d_0},
\label{delta_beta3d}
\end{equation}
where $L_0$ is a constant used to set the scale. We chose $L_0=20$,
which gives on a $20\times 20$ lattice the same step-size as we used 
in~I. The volume independent step-size (\ref{step_size}) makes one 
sweep a physical time unit for model~A type fluctuations in nature. 
It has the nice side effects that it allows to simulate larger volume 
lattices, but questions about the survival of effects in the limit of 
a slow dynamics become more difficult to answer.

We build ensembles of at least 160 cycles, grouped in bins of 20 cycles 
each. The error bars are obtained with respect to 32 jackknife bins for 
smaller lattices and 8 for larger lattices. The system is driven between 
$\beta_{\min}=0.2$ and $\beta_{\max}=0.4$. It is equilibrated in the 
disordered phase for at least $80$ sweeps at $\beta_{\rm min}$, before 
being driven through a hysteresis cycle.

\subsubsection{Latent Heat\label{hsec_e} }

As in I we use the maximum opening of the energy hysteresis as estimator
for the latent heat on a finite lattice (some energy hysteresis pictures
are given in~I). Now we perform the limit $L\to\infty$ for a fixed 
dynamics defined by $n'_{\beta}$. The dynamics is at a constant speed in
physical units, in contrast to the $n_{\beta}$ dynamics used in I, which
becomes infinitely slow in the limit of large lattices.

\begin{figure}[ht] \vspace{-2mm} \begin{center}
\epsfig{figure=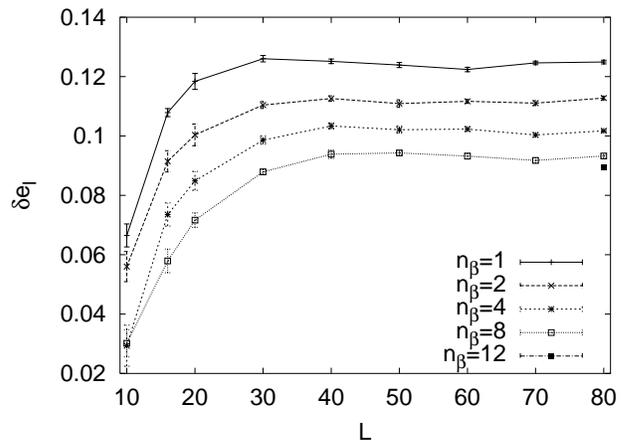,width=\columnwidth} 
\vspace{-1mm}
\caption{The 3D 3-state Potts model finite size estimates of the 
latent heat for various $n'_\beta$ dynamics at zero external magnetic 
field.} \label{hys_lh3d}
\end{center} \vspace{-3mm} \end{figure}

Relying on energy hysteresis curves for the 3D $3$-state Potts model at 
zero external magnetic field, we plot in Fig.~\ref{hys_lh3d} the 
lattice size dependence of their maximum openings. A plateau is reached 
for $L\ge 30$ for the $n'_\beta=1$ dynamics and for $L\ge40$ for the 
large values of $n'_\beta=4$ (slower dynamics). The error bars given in 
Fig.~\ref{hys_lh3d} are purely statistical and do not take into account 
a possible systematic error, which comes from the $L$ independent 
constant step-size of the $n'_{\beta}$ dynamics~\cite{s-error}. 

As plateaus for $\triangle \overline{e_l}(n'_\beta,L)$ are reached at 
$L\ge 40$, it is save to take the $L=80$ latent heat values as the our 
final answers. They are listed in the following.
$n'_{\beta}=1$: $0.1249\,(7)$,
$n'_{\beta}=2$: $0.1128\,(7)$,
$n'_{\beta}=4$: $0.1017\,(5)$, 
$n'_{\beta}=8$: $0.0932\,(6)$, and
$n'_{\beta}=8$: $0.0895\,(6)$.
They are much larger than the equilibrium estimate 
$\triangle \overline{e_l}=0.05354\,(17)$ \cite{Al91}. The slower speed 
of the transition at higher $n'_\beta$ lowers the values, but in the 
limit $n'_{\beta}\to \infty$ we still expect a dynamical latent heat 
which is larger than the equilibrium result. Namely, the $L\to\infty$ 
limit of the dynamics of~I provides a lower bound and a dynamically 
generated latent heat was observed there for 2D Potts models, including
the Ising model. In 3D one may want to modify the dynamics which gives a 
lower bound from $\triangle\beta \sim 1/V$ of~I to $\triangle\beta\sim 
1/L^2$, because the slowing down at the critical point of the Ising
model scales with $L^z$ and $z\approx 2$ (see Ref.~\cite{LBbook} for 
estimates of $z$). Fitting our present 
$\triangle\overline{e_l}(n'_\beta)$ data to the form 
$\triangle \overline{e_l}(n'_\beta) = \triangle \overline{e_l}
+ a\,(n'_{\beta})^b$ gives $\triangle \overline{e_l}=0.057\,(9)$
with an exponent $b=-0.298\,(55)$.
The rather large error bars prevent a conclusion about whether there
is a dynamically generated latent heat in 3D or not. A change of the
dynamics and/or a push towards larger $n'_{\beta}$ values is required. 

\subsubsection{Structure Functions\label{Hsec_sf}}

\begin{figure}[ht] \vspace{-2mm} \begin{center}
\epsfig{figure=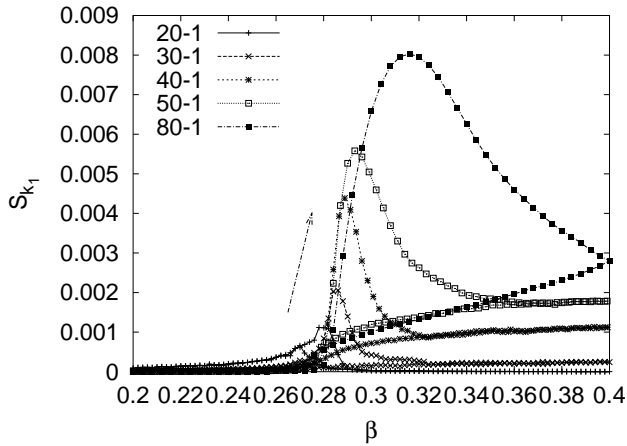,width=\columnwidth} \vspace{-1mm}
\caption{Hysteresis of the structure function $S_{k_1}(\beta)$ on
various $L^3$ lattices for the 3D 3-state Potts model in zero external 
magnetic field.} \label{fig_hys_sf1k3d}
\end{center} \vspace{-3mm} \end{figure}

\begin{figure}[ht] \vspace{-2mm} \begin{center}
\epsfig{figure=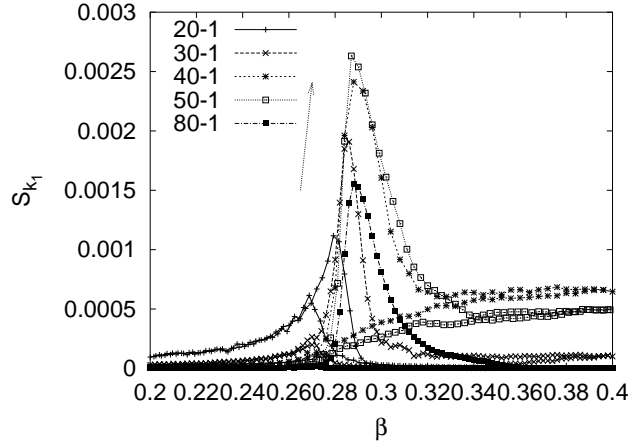,width=\columnwidth} \vspace{-1mm}
\caption{Hysteresis of the structure function $S_{k_1}(\beta)$ on
various $L^3$ lattices for the 3D 3-state Potts model in the external 
magnetic field $h=0.0005$.} \label{fig_hys_sf1k3dh}
\end{center} \vspace{-3mm} \end{figure}

The structure function behavior is similar to the 2D case with 
pronounced peaks in the cooling half-cycle. In Fig.~\ref{fig_hys_sf1k3d} 
we plot $S_{k_1}$ for the $n'_\beta=1$ dynamics of the 3D 3-state Potts 
model in zero magnetic field. In Fig.~\ref{fig_hys_sf1k3dh} we show 
$S_{k_1}$ in the external magnetic field $h=0.0005$, which is slightly 
higher than the critical end-point field. Recall that $k_i$ depends
on the lattice size $L$ through Eq.~(\ref{momenta}).

Arrows indicate the flow with increasing $\beta$. At $\beta=0.4$ the 
hysteresis turns around and the lower curves describe the $\beta_{\max}
\to \beta_{\min}$ approach. For the smaller lattices (most clearly for
$L=20$) they exhibit small peaks around the transition temperature, 
which are similar in size to peaks one finds for equilibrated 
configurations. For the larger lattices these peaks are not seen. 
Their returning curves come in too high, because their hysteresis are 
at $\beta=0.4$ still far out of equilibrium. 

As expected, the $\beta_{\min}\to\beta_{\max}$ peaks are smaller in the 
presence of the external magnetic field. While for $h=0$ they increase 
over the entire range of our lattices, we find for $h=0.0005$ an initial 
increase, which turns around between $L=40$ and 60. However, the 
subsequent decrease appears to be slow compared to non-critical behavior 
for which the structure functions $S_{k_i}$ falls off $\sim 1/V$ in our 
normalization. Possibly, the behavior is similar to the equilibrium 
behavior at a second order phase transition point, where the fall-off 
is $\sim 1/L^x$ and $0<x<3$. 

\begin{figure}[ht] \vspace{-2mm} \begin{center}
\epsfig{figure=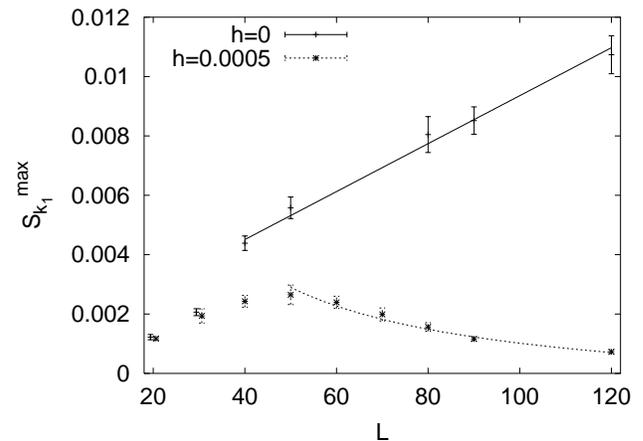,width=\columnwidth} \vspace{-1mm}
\caption{Structure function $S_{k_1}$ maxima versus lattice size
for the $n'_{\beta}=1$ dynamics of the 3D 3-state Potts without 
external magnetic field ($h=0$) and in the cross-over region
($h=0.0005$). For $L=20$ and 30 the data points are slightly 
shifted, so that it becomes visible that there are results for 
both $h=0$ and $h=0.0005$.} \label{fig_sfmax}
\end{center} \vspace{-3mm} \end{figure}

Including lattices as large as $120^3$, we plot in Fig.~\ref{fig_sfmax} 
our estimates of the structure function maxima for $S_{k_1}$ ($h=0$ and 
$h=0.0005$) together with fits for the larger lattices. Unfortunately, 
the precision of the data does not allow to determine the exponent $x$ 
in a fit of the form 
\begin{equation} \label{Skmax_fit}
S_k^{\max} = a_1 +a_2\,L^x
\end{equation}
accurately. Choosing the exponent by hand, $x=+1$ for $h=0$ and $x=-1$ 
for $h=0.0005$, we find a satisfactory goodness-of-fit $Q$ (for the
definition of $Q$ see, e.g., Ref.~\cite{NumR}) in each case. Using the 
non-critical $x=-3$ (instead of $x=-1$) to fit the $h=0.0005$ data, 
gives the too small goodness-of-fit $Q=0.0062$. Of course, this cannot 
exclude that the behavior turns to non-critical for even larger 
lattices. In this connection it is puzzling that the value $x=+1$ 
used to fit the $h=0$ data cannot be asymptotic. In our normalization 
the $S_k$ structure functions are bounded from above by $S_k\le 1$, 
implying $x\le 0$ for any true asymptotic behavior. Apparently lattices 
of size $L>120$ are needed. With checkerboard updating they could be 
handled on supercomputers .

\begin{figure}[ht] \vspace{-2mm} \begin{center}
\epsfig{figure=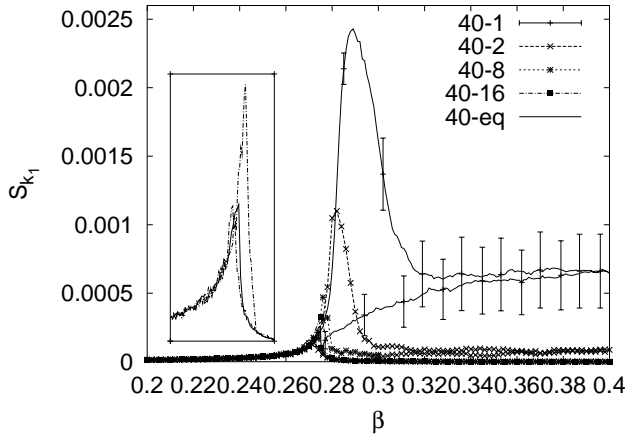,width=\columnwidth} 
\vspace{-1mm}
\caption{The structure function $S_{k_1}$ hysteresis for the 3D 3-state 
Potts model in the $h=0.0005$ external magnetic field on a $40^3$ 
lattice for simulations with the $n'_\beta$ dynamics indicated by the 
extension to the lattice size. The inlay enlarges the equilibrium peak 
together with the slowest dynamics $n_\beta=16$ data ($\beta$ is mapped 
on $0.21+2\,(\beta-0.26)$ and $S_{k_1}$ on $4\,S_{k_1}+0.0001$).}
\label{fig_hys_sf1kV3dh}
\end{center} \vspace{-3mm} \end{figure}

For a $40^3$ lattice we plot in Fig.~\ref{fig_hys_sf1kV3dh} the 
structure function $S_{k_1}$ for the 3D 3-state Potts model in the 
$h=0.0005$ field for different $n_\beta$ up to $16$ together with 
equilibrium data. 
The approach to equilibrium with the decrease of the speed of the 
dynamics (increase of $n'_{\beta}$) appears to be faster than in 2D.
Possibly a greater connectedness of domains allows in higher dimensional 
models for a faster evolution of macroscopic structures. 
The heating ($\beta_{\max}\to\beta_{\min}$) peak develops and reaches 
the equilibrium peak, as we reduce the speed. The 
cooling peak decreases, but it would need much slower simulations 
to reach the equilibrium values. Thus within a range of speeds the 
cooling peak stays strong and spinodal decomposition is the dominant 
scenario. For $h=0$ the figure looks very similar, but for an increase 
of all peak values by a factor of about two. The explanation of the 
difference between the heating and cooling branches presented for 2D 
models in~I remains valid. Driven from the disordered to the ordered 
phase, the system freezes in domains of different magnetization, which 
are slow to evolve. For a fast dynamics it does not have enough time 
to equilibrate in the ordered phase. The decrease of the speed allows 
then the system to equilibrate in the ordered phase and the heating 
peak emerges.

\begin{figure}[ht] \vspace{-2mm} \begin{center}
\epsfig{figure=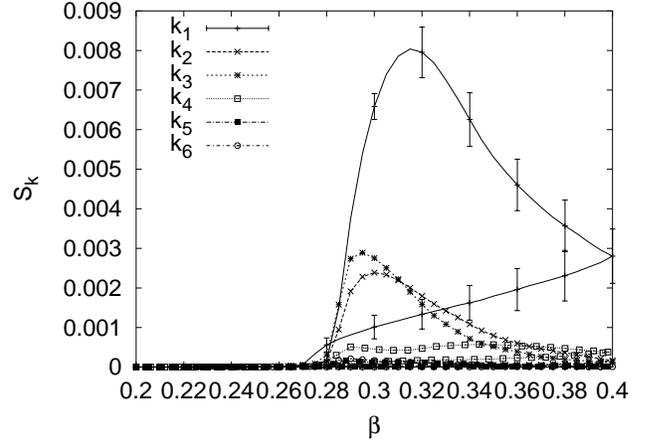,width=\columnwidth} 
\vspace{-1mm}
\caption{Hysteresis for the structure functions for the 3D 3-state 
Potts model on an $80^3$ lattice in zero external field.}
\label{fig_hys_hsfk3d}
\end{center} \vspace{-3mm} \end{figure}

\begin{figure}[ht] \vspace{-2mm} \begin{center}
\epsfig{figure=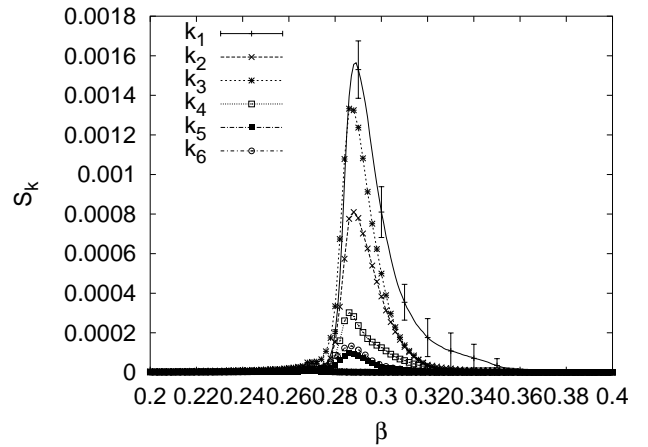,width=\columnwidth} 
\vspace{-1mm}
\caption{Hysteresis for the structure functions for the 3D 3-state 
Potts model on an $80^3$ lattice in the $h=0.0005$ external field.}
\label{fig_hys_hsfk3dh}
\end{center} \vspace{-3mm} \end{figure}

In Fig.~\ref{fig_hys_hsfk3d} and~\ref{fig_hys_hsfk3dh} we plot for an
$L=80$ lattice the lowest momentum ($k_1,...,k_6$) structure functions 
for zero and non-zero external magnetic field. In the absence of the 
external magnetic field we observe that the peaks are very pronounced
for $k_1$ to $k_3$, and much smaller for $k\ge k_4$. In the presence of 
the external magnetic field all peaks are smaller and the difference 
between lower and higher modes is less distinct. We still observe a 
qualitative difference between the $k\le k_3$ and the $k\ge k_4$ modes. 
Another observation is that the peaks are less broad for $h=0.0005$ 
than for $h=0$, because the system equilibrates for $h>0$ easier in 
the ordered phase.

\subsubsection{FK Clusters\label{Hsec_FK}}

\begin{figure}[ht] \vspace{-2mm} \begin{center}
\epsfig{figure=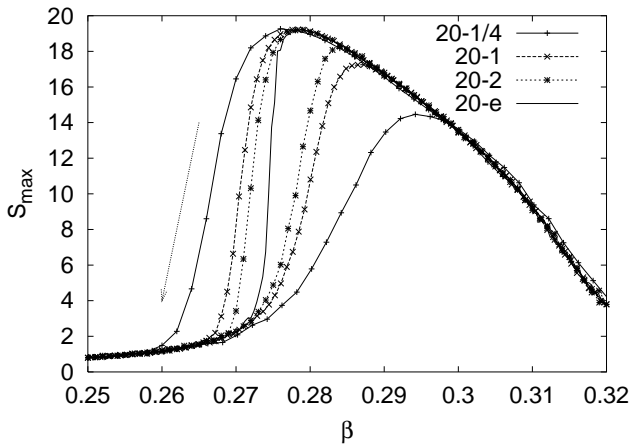,width=\columnwidth} \vspace{-1mm}
\caption{Hysteresis for the largest connected cluster surface 
$S_{\rm max}$ for the 3D 3-state Potts model in the external magnetic 
field $h=0.0005$ on a $20^3$ lattice. The hysteresis flow is indicated
by the arrow. The $n'_\beta$ values are given the extensions to the 
lattice size and $e$ stands for equilibrium.} \label{hys_clsmVh}
\end{center} \vspace{-3mm} \end{figure}

In~I we identified the maximum connected surface of FK clusters as an
interesting quantity. In the transition region it exhibits a pronounced 
peak related to percolation. The 3D analysis \cite{Vel} of this quantity 
has remained limited to rather small lattices ($L\le 30$), because of a 
rapid slowing down with lattice size of the algorithm identifying the 
maximum connected surface. In contrast to 2D deviations from equilibrium 
appear relatively weak. A reason may be
that the extra dimension adds new degrees of freedom to the surface.
For the crossover region ($h=0.0005$) the approach to equilibrium is 
illustrated in Fig.~\ref{hys_clsmVh} by comparing the dynamics at 
different speeds (including a `super-fast' $n'_{\beta}=1/4$ choice) with 
equilibrated data. Notably the figure is practically identical for 
the $h=0$ transition, reiterating that percolation happens for a 
crossover in quite the same way as for a proper transition~\cite{Sa01}.

\section{Summary and Conclusions\label{sec_conclude}}

In our simulations of the model~A dynamics of the 3D 3-state Potts 
model on $L^3$ lattices, we find spinodal decomposition to be the 
dominant scenario of phase conversion. Under a quench from the 
disordered into the ordered phase, we observe an early time 
development of structure functions $S_k(t)$, which is in over-all
agreement with the exponential growth predicted by the linear theory 
of spinodal decomposition for $k<k_c$ (Fig.~\ref{fig_omega_3dsq}). 
However, the ordering of the growth coefficients is not always in 
agreement with Eq.~(\ref{omega_k}). Further, the linear approximation 
fails for near critical quenches (as was known before~\cite{La92}) 
and, altogether, in 2D. 

Following the structure function evolution over extended times, their 
pronounced peaks are the most noticeable feature. It is certainly a 
challenge to base a theory of spinodal 
decomposition on their finite size scaling instead of using the 
linear approximation. Unfortunately, very large lattices appear
to be needed to extrapolate the infinite volume limit for the
structure function maxima. Within our hysteresis investigation we 
pushed their analysis to systems as large as $120^3$. But, for zero 
external magnetic field we find an increase, which cannot yet be 
asymptotic, because it violates an upper bound. Applying a small 
external magnetic field ($h=0.0005$) and quenching  through the 
cross-over region results in an increase with $L$ for small lattices 
($L\le 40$) and a {\em slower than non-critical} fall off on our 
larger ($L\ge 50$) lattices (Fig.~\ref{fig_sfmax}). If the last
trend continues towards $L\to\infty$, signals of spinodal decomposition
would survive the continuum limit of an analogue QCD scenario.

A quench into the ordered phase leads to a competition of FK 
cluster domains. For the proper transition ($h=0$) this leads to 
a divergence of the equilibration time in the limit $L\to\infty$,
an effect well-known in condensed matter physics~\cite{ChLu_equi}. 
For a cross-over the $L\to\infty$ scenario is less scary, but our 
Fig.~\ref{quench:3qh} shows that one may still expect a substantial 
time delay before reaching equilibrium.
In the RHIC experiment one performs a near-critical quench into the
quark-gluon plasma phase (assuming the phase is reached). For a 
near-critical quench FK clusters are initially small. In our 
Fig.~\ref{quench:3q_diff} the largest one covers only about 10\% 
the system. So, a number of vacuum domains are expected 
and the effects from the dynamics are manifested on their surfaces. 
Inside the domains the system may equilibrate fast, but the 
system as a whole cannot reach equilibrium easily. Unfortunately,
a concise definition of FK clusters is not available for lattice
gauge theory. Therefore, one may turn towards investigating 
directly the SU(3) gluonic energy density for influences of the 
supposedly underlying vacuum structure. First results will be
published in Ref.~\cite{BBV}.
\medskip

\acknowledgments
BB and AV would like to thank Urs Heller and Michael Ogilvie for useful 
discussions. This work was in part supported by the US Department of 
Energy under contract DE-FG02-97ER41022. The simulations were performed 
on PC clusters at FSU and IUB.


\begin{thebibliography}{19}
\bibitem{cooling} As in~I we use the spin model notation for the 
temperature. Our cooling quench corresponds to the heating quench
in laboratory deconfining experiments.
\bibitem{s-error} The constant step-size introduces slight problems in 
determining the precise location of the maximum (which in our dynamics 
of~I went away for large $L$). This difficulty may lead to systematic 
errors, which could be overcome by using the slower random updating 
and taking data within a sweep. We did not do so as the systematic 
uncertainty appears to be smaller than the statistical. In particular,
it decreases for increasing $n'_{\beta}$.
\end{thebibliography}
\end{document}